\documentclass[10pt,letterpaper]{article}
\usepackage[OE]{express}
\usepackage{amssymb}
\usepackage{wasysym}  
\usepackage{graphicx}
\usepackage{multirow}
\usepackage{color}
\usepackage{adjustbox}
\usepackage{slashbox}

\begin{document}

\def\crta{\vrule height1.41ex depth-1.27ex width0.34em}
\def\dj{d\kern-0.36em\crta}
\def\Crta{\vrule height1ex depth-0.86ex width0.4em}
\def\Dj{D\kern-0.73em\Crta\kern0.33em}
\dimen0=\hsize \dimen1=\hsize \advance\dimen1 by 40pt

\title{Mixed basis quantum key distribution with linear optics}

\author{Mladen Pavi{\v c}i{\'c},\authormark{1,2,5} 
Oliver Benson,\authormark{1,6}
Andreas W.~Schell,\authormark{3} and Janik Wolters\authormark{4}}

\address{\authormark{1}Department of Physics, Nanooptics, 
Math.-Nat. Fakult\"at, Humboldt-Universit\"at zu Berlin, Germany\\
\authormark{2}Center of Excellence for Advanced Materials 
and Sensing Devices (CEMS), Photonics and Quantum Optics Unit, 
Ru{\dj}er Bo\v skovi\'c Institute, Zagreb, Croatia\\
\authormark{3}Dep. of Electronic Science \&\ Engineering, 
Kyoto Univ., Kyoto daigaku-katsura, Nishikyo-ku, Kyoto, Japan\\
\authormark{4}Department of Physics, University of Basel, 
Switzerland\\
\authormark{5}mpavicic@physik.hu-berlin.de\\
\authormark{6}oliver.benson@physik.hu-berlin.de}


\begin{abstract}
Two-qubit quantum codes have been suggested to obtain better 
efficiency and higher loss tolerance in quantum key distribution. 
Here, we propose a two-qubit quantum key distribution protocol 
based on a mixed basis consisting of two Bell states and two states 
from the computational basis.  All states can be generated from a 
single entangled photon pair resource by using local operations 
on only one auxiliary photon. Compared to other schemes it is also 
possible to deterministically discriminate all states using linear 
optics. Additionally, our protocol can be implemented with today's 
technology. When discussing the security of our protocol we find a 
much improved resistance against certain attacks as compared to the 
standard BB84 protocol.
\end{abstract}



\bibliographystyle{osajnl}




\section{Introduction}
\label{sec:intro}

Quantum key distribution (QKD) promises secure information 
transfer based on the laws of quantum physics.
The most prominent protocol is the famous BB84 protocol \cite{bb84}. 
It is proven to be {\em unconditionally\/} secure provided 
that Alice and Bob make use of a genuine random 
number generator \cite{stipcevic-ursin-15} and that the quantum 
bit error rate (QBER) is below 11\%\ \cite{gisin-02}. The latter can
be increased to 12.7\%\ for the six-state protocol \cite{lo-01a}. 
With a two-way classical communications QBER can be increased 
further and we shall come back to this point in 
Sec.~\ref{sec:conclusion}. The secure QBER can also be increased
by increasing the capacity of the protocol so as to send 3 or
4 messages (in contrast to 2 in BB84) via three or four states 
in a 3- or 4-dim space, achieving 22.7\%\ or 25\%\, respectively. 
\cite{bech-00,bruss-02}.

Beyond maximum tolerable QBER Eve (an eavesdropper) is undetectable 
when the losses are significant, however, she might be undetectable 
even below that limit because weak laser pulses, standardly used for 
implementing QKD with single photons, enable her a beam-splitting 
\cite[8.5.3, p.~440]{dusek-lutkenhaus-06} and a photon-number 
splitting attack \cite[8.5.4, p.~441]{dusek-lutkenhaus-06}. Also, 
recent commercial systems based on BB84 protocol were shown to be 
hackable by tailored bright illumination and that initiated 
``identifying and patching technological deficiencies'' of BB84 
implementation \cite{lydersen-10}.

Therefore, time and again over the last decade a QKD with entangled
photons has been considered and reconsidered mostly as modifications
of the so-called {\em ping pong\/} (pp) protocol with two Bell states
\cite{bostrom-felbinger-02,bostrom-felbinger-08}, 
by means of all four Bell states, i.e. via the superdense coding (SDC)
protocol \cite{cai-li-04}, or even with three particle GHZ state
\cite{chamoli-09}. It can be argued that ``the potential
of entanglement-based protocols need to be seriously explored,
especially taking into account the rapid research progress 
\cite{ngah-15} of entanglement light sources" \cite{chen-16}. 
On the other hand, correlated detections of photons from the
same down-converted pairs provide us with higher loss tolerance. 

Various kinds of attack on these deterministic protocols were
designed \cite{wojcik-03,bostrom-felbinger-08,pavicic-pra-13} and
to cope with them a number of modifications of the protocols were
put forward \cite{nguyen-04,yang-11,zawadzki-15,li-jin-13,han-14}.
Several implementations have been carried out
\cite{ostermeyer-08,chen-16} and the security considered
\cite{wojcik-03,nguyen-04,bostrom-felbinger-08,yang-11,zawadzki-15,li-jin-13,han-14}
although an unconditional one has not been reached. 

One of the first proposed attacks on the pp protocol, given by Nguyen
\cite{nguyen-04,man-05}, enables Eve to read all the messages in the
message mode ``absolutely unnoticeable.'' This kind of attack has 
been addressed in \cite{han-14,chen-16} and the protocol shown to be
secure via its control mode. Nguyen's pp modification, called 
{\it quantum dialog\/} (in which both, Alice and Bob, send entangled 
photons and messages) has been addressed in \cite{man-05}. 

As we show in Sec.~\ref{sec:security}, Nguyen's attack can be easily
extended to the aforementioned four-Bell-state pp protocol which was
proposed to increase the capacity of the protocol---by a transfer of
2 bits via 4 messages---but which requires non-linear optics elements
\cite{kim-kulik-shi-01} (it cannot be carried out with linear optics
ones \cite{vaidman99,luetkenhaus-99}).

In this paper we propose a high-capacity (four messages)
entanglement-based pp-like protocol which is not only resistant 
to Nguyen's attack but also enables Alice and Bob to detect Eve 
during their data exchange without switching to a separate control 
mode and which can be implemented with linear optics elements 
because it is based on the mixed basis consisting of two Bell 
states and two states from the computational basis.

The paper is organized as follows. In Sec.~\ref{sec:setup} we
introduce the basis our states are in and a description of our
setup. In Sec.~\ref{sec:protocol} we outline our protocol.
In Sec.~\ref{sec:security} we discuss the security of the
protocol and in Sec.~\ref{sec:conclusion} we summarize the
results we achieved.

\section{Mixed basis and setup}
\label{sec:setup}

Let us start with introducing the mixed basis used in our protocol.
We define it as a basis consisting of the 
 two Bell states
\begin{eqnarray}
|\chi^{1,2}\rangle=|\Psi^\mp\rangle=\frac{1}{\sqrt{2}}(|H\rangle_1|V
\rangle_2\mp|V\rangle_1|H\rangle_2)
\label{eq:bell-states} 
\end{eqnarray}
and the two computational basis states
\begin{eqnarray}
|\chi^3\rangle=|H\rangle_1|H\rangle_2, 
\qquad |\chi^4\rangle=|V\rangle_1|V\rangle_2,
\label{eq:com-states} 
\end{eqnarray}
where $|H\rangle_i$ ($|V\rangle_i$) represents a 
horizontal (vertical) polarized photon in mode $i$.  

\begin{figure}[htbp]
\centering
\includegraphics[width=0.85\textwidth]{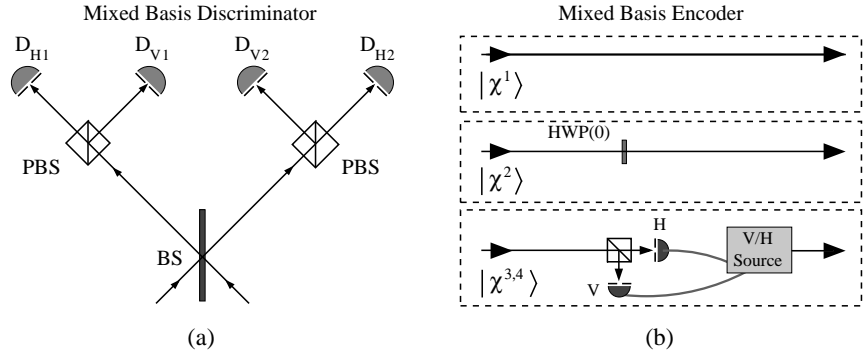}
\caption{\label{fig:analyser}(a) {\em Setup for the mixed 
basis state discrimination.\/}
All four states of the mixed basis can be deterministically 
discriminated by using only a few linear optic 
elements like a beam splitter (BS), two polarizing 
beam splitters (PBS), and photon number resolving detectors (D). 
(b) {\em Setups for mixed basis encoding.\/} All 
four mixed basis states can be generated from $\Psi^-$ by 
manipulating \emph{one} of the two photons with one of the 
shown devices (see text). While for $\chi^{1,2}$ the state preparation is 
fully deterministic, it is heralded for 
$\chi^{3,4}$.}
\end{figure}

A particular advantage of the mixed basis is that all 
four basis states can be deterministically discriminated 
using only a few linear optical elements. 
The discrimination setup consists of a non polarizing beam 
splitter (BS) and an additional polarizing beam splitter (PBS) 
in each of its two output ports---see Fig.~\ref{fig:analyser}.

The outputs of the PBSs are monitored by four photon 
number resolving detectors 
\cite{s-phot-det05,thomas-10,fukuda-11}. They can 
also be approximated by purely linear elements such as 
additional concatenated beam splitters and single photon 
detectors \cite{mattle-zeil-96,hadfield-09,lita-nam-08}.

In the case of 
$|\chi^3\rangle=|H\rangle_1 |H\rangle_2$ and 
$|\chi^4\rangle=|V\rangle_1 |V\rangle_2$ as input states of 
the discriminator, two indistinguishable parallel polarized 
photons are sent to the BS from different sides, as shown in 
Fig.~\ref{fig:analyser}(a). These photons will always exit the beam 
splitter at the same side, bunched together and showing the 
well known Hong-Ou-Mandel interference effect~\cite{ou-book07}.
Both bunched photons keep the polarization direction they had 
before they entered the beam 
splitter~\cite{p-pra94,p-s94,pavicic-book-05}.
On the other hand, $|\chi^{2}\rangle=|\Psi^+\rangle$ 
photons bunch together behind the BS, but have different
polarization and split at a PBS
behind the BS. In contrast, $|\chi^{1}\rangle=|\Psi^-\rangle$ 
photons split at the BS and are subsequently transformed into 
opposite polarization states by the PBSs.
Thus, all four mixed basis states can be deterministically 
and unambiguously discriminated by means of photon 
number resolving detectors. 

\begin{figure}[htbp]
\centering
\includegraphics[width=0.7\textwidth]{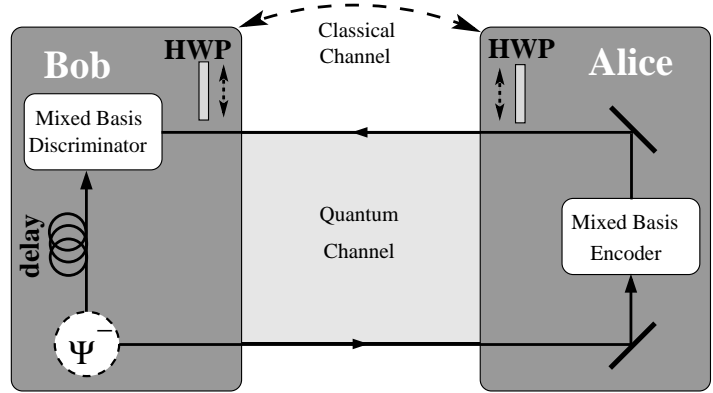}
\caption{\label{fig:bs-bs}{\em QKD setup.\/} Bob's part 
consists of an entangled photon pair source generating the state $\Psi^{-}$, 
a quantum delay, a mixed basis discriminator and a removable HWP($\frac{\pi}{8}$) 
aligned to $\pi/8$. Alice's part consists of a mixed basis encoder and a  
removable HWP($\frac{\pi}{8}$) also aligned to $\pi/8$. Alice and Bob exchange 
information on the bases and states over a classical channel.}
\end{figure}

In the following, we demonstrate how to prepare the mixed basis 
states solely from the state $|\Psi^-\rangle$ generated by an 
entangled photon source \cite{mattle-zeil-96,jin-14}.
In order to do this we introduce also our QKD setup in
Fig.~\ref{fig:bs-bs}, which consists of Alice's and Bob's part
together with a quantum and classical communication channel. Bob
has an entangled photon source, a quantum delay, and a mixed basis 
discriminator, as shown in Fig.~\ref{fig:analyser}(a).
We assume that Bob's entangled photon source generates photon pairs 
in state $|\chi^1\rangle=|\Psi^-\rangle$. Bob sends one of the photons
of the pair, the {\em travel\/} photon, to Alice and keeps the other, 
the {\em home\/} photon, delayed for later joint measurements with the
travel photon returning from Alice. 

Now, Alice can prepare any state of the mixed basis semi-deterministically 
by manipulating the travel photon she receives from Bob.
For this she has a mixed basis encoder, which consists of an auxiliary 
on-demand single photon source and linear optical elements as shown in 
Fig.~\ref{fig:analyser} (b). We will first consider the case when 
no additional HWP($\frac{\pi}{8}$) on Bob's or Alice's side, 
shown in Fig.~\ref{fig:bs-bs}, are put in. 

To generate $|\chi^1\rangle=|\Psi^-\rangle$ nothing need be 
done---Fig.~\ref{fig:analyser}(b), top). To generate 
$|\chi^2\rangle=|\Psi^+\rangle$  Alice puts a half-wave plate 
[${\rm HWP}(0^\circ)$] into the photon 
path---Fig.~\ref{fig:analyser}(b), middle row. 
The HWP changes the sign of the 
vertical polarization and thereby transforms 
$|\Psi^-\rangle$ into $|\chi^2\rangle=|\Psi^+\rangle$. 
Thus, $|\chi^{1,2}\rangle$ can be generated deterministically.

To generate  $|\chi^{3,4}\rangle$ Alice places a PBS with single 
photon detectors on its output ports and an auxiliary single photon 
source in her photon's path---Fig.~\ref{fig:analyser}(b), lower row. 
With this, the polarization of the photon is measured and thereby 
Bob's home photon is projected into $|V\rangle$ if Alice's measurement 
of the travel photon gives  $|H\rangle$ and into $|H\rangle$  
if Alice measured $|V\rangle$. Subsequently, Alice replaces 
the destructively measured photon with a photon of opposite 
polarization from an auxiliary single photon source. In doing so, 
it is possible to generate 
$|\chi^3\rangle=|H\rangle_1 |H\rangle_2$ and
$|\chi^4\rangle=|V\rangle_1 |V\rangle_2$ 
in a heralded way. State generation of $|\chi^{3,4}\rangle$ is 
hence probabilistic in the sense that Alice obtains 
$|H\rangle$ or $|V\rangle$ after her PBS completely at random, 
but as soon as she does obtain them, the $|\chi^{3,4}\rangle$
states are determined because she knows what
she sent and Bob will measure. 

Let us now come back to the HWP($\frac{\pi}{8}$) shown in  
Fig.~\ref{fig:bs-bs}. 
We assume that Alice and Bob, independently of 
each other and randomly insert their HWPs aligned to $\pi/8$ 
(Hadamard gates) using a true quantum random number generator 
\cite{stipcevic-ursin-15}.

When both HWPs are inserted Bob will receive (in the absence 
of Eve) the same states as with no HWP inserted (since two 
consecutive Hadamards yield the identity). These two 
arrangements we call the {\em same bases\/}. 
The case when only Alice's or only Bob's HWPs are inserted 
we call {\em different bases\/}.

With a delay Bob informs Alice of his choice of bases and
Alice him of hers, over a classical channel. The data obtained
with different bases serve them as {\em control data\/} which
they use to catch Eve. Alice, also with a delay, informs Bob
of exact values of all control data. Since the capacity of the
classical channels is practically unlimited compared to the
quantum channel, the quantity of classical information exchanged
should not be a problem. Handling of messages obtained with
different bases and the corresponding control data 
we call the {\em control mode\/}.

\section{Protocol}
\label{sec:protocol}

Now we describe how Alice and Bob using the setup from 
Fig.~\ref{fig:bs-bs} proceed to securely exchange a one-time-key. 
Our suggested protocol is the following:

1. Bob decides randomly about his basis by placing or
  not placing his HWP. He prepares the two photon state
  $|\chi^1\rangle$, stores the \emph{home photon} in his
  quantum delay, and sends the \emph{travel photon} to
  Alice through a quantum channel. 

2. Alice decides randomly about her basis by placing or not 
placing her HWP. She decides to perform the coding operation 
for $|\chi^1\rangle$, $|\chi^2\rangle$ or 
$|\chi^{3,4}\rangle$ (with genuine quantum randomness) 
with probabilities $P(\chi^{1-4})=1/4$.

3. Bob measures the two qubit state by means of his mixed
  basis discriminator, and broadcasts on a public channel if
  he placed his  HWP or not.

4. Alice checks if the transmitted message is valid. This is
  the case for all $|\chi^{1-4}\rangle$ messages if the bases were
  the same. If a message is valid, it is stored for later usage
  as one time key. The selection is called {\em sifting\/}. 
  If the bases are different, the
  measurement data are stored for later usage as the control data.
  Alice announces the valid messages via a public channel with a
  delay. She also announces the values of all control data. 

5. Bob repeatedly restarts with (a) and with a delay he
  processes the control data to check for Eve's presence.
  If yes, they abort the transmission. If not, they distill the
  key. Thereupon they carry out error correction and privacy
  amplification \cite{gisin-02}.

We would like to stress that even in the case of different bases 
Alice's and Bob's measurements are sensitive to detect the 
eavesdropper Eve. This is different from the BB84 protocol where 
part of the valid data has to be sacrificed. 
Also no active switching between a message and control mode has 
to be performed as in other pp-like protocols.
This is an advantage since switching may open the possibility to 
advanced eavesdropping attacks, in particular when Eve can hide 
her presence in the message mode completely as shown below. 

\section{Security}
\label{sec:security}

As we mentioned in Sec.~\ref{sec:intro}, Nguyen's attack
\cite[p.~7, par.~containing Eq.~(2)]{nguyen-04} is a
powerful attack on pp-like protocols. We start with its brief 
presentation and only then we show that our protocol is 
resistant to it. 

\begin{figure}[htbp]
\centering
\includegraphics[width=0.99
\textwidth]{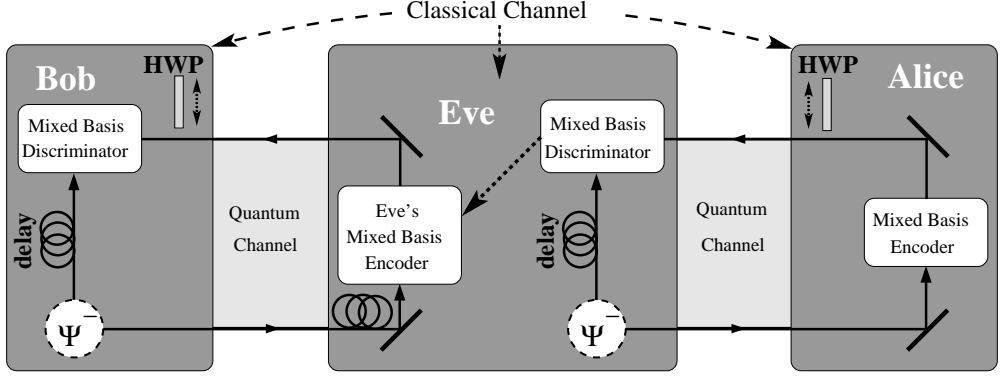}
\caption{\label{fig:bs-eve-bs}{\em An efficient attack by Eve.\/} 
Eve delays Bob's photon and sends a photon from her own 
$\Psi^-$  ($\chi^1$) source. Alice prepares messages for 
Bob, but Eve deterministically reads them off but can only 
probabilistically prepare states of Bob's home photons. 
For instance, when detecting that Alice prepared  
$|\chi^3\rangle$ ($|HH\rangle$) she can replicate it only 
with a 50:50 probability, because Bob's home photon is  
entangled with the travel photon handled by Eve. Nguyen's attack
\cite[p.~7, par.~containing Eq.~(2)]{nguyen-04} on pp can be viewed
as the one shown here only without HWPs and without the classical
channel.}   
\end{figure}

As shown in Fig.~\ref{fig:bs-eve-bs}, Eve delays the 
photons Bob sent to Alice and instead, sends her own
photons from her $\Psi^-$ source to Alice to encode 
them. Eve intercepts the photons Alice encoded, measures 
them in her discriminator, encodes the read messages on 
the travel photons she kept delayed and sends them to Bob. 
Eve can read off all the states sent by Alice in the p-p
protocol but cannot in the protocol of ours. 

Our protocol is less susceptible to the above attack because:

1. When Alice prepares $|\chi^{3,4}\rangle$ states she collapses 
the states of both photons -- her and Eve's. Eve can 
deterministically find out which states Alice's and her photons 
collapsed to, but can collapse Bob's photon states into ones of her 
own photons only with a probability of 50\%\ (by means of her HWP);  

2. When Alice and Bob put HWPs in their channel, then
  both Eve's reading and resending will be scrambled. 
 
According to \cite{fuchs-gisin-peres-97,bruss-02,scarani-09} we 
evaluate the condition for QKD to be secure in the presence of Eve: 
$I_{AB}>I_{AE}$, where $I_{AB}$ ($I_{AE}$) is the mutual information 
between Alice and Bob (Eve). We calculate them as follows.

When Alice and Bob are in the same basis with no HWPs inserted and
Alice sends $|\chi^1\rangle$ or $|\chi^2\rangle$ Eve can detect them
and impose the same state on the Bob's photons. Eve is not necessarily
always present in the line and we shall denote her presence by
$X\in [0,1]$; $X=0$ means that Eve is not present at all and $X=1$
that she is always present. Thus, Bob will always receive the correct
$|\chi^1\rangle$ or $|\chi^2\rangle$, each with the probability $1/4$,
no matter whether Eve is present or not (when she is in the line she
fatefully transmits what she reads), but Eve will read
$|\chi^{1,2}\rangle$ only with the probability $X/4$, i.e., only when
she is in the line. When Alice sends $|\chi^3\rangle$  or
$|\chi^4\rangle$ Eve can detect them but cannot impose the same
state on the Bob's photons with a probability higher than 50\%.
So, Bob's probability of receiving a correct state via Eve diminishes
with her presence $X$ (probability falls as $(2-X)/8$) and of receiving
incorrect state increases with $X$. Eve's probability of receiving
both correct and incorrect states increases with $X$. 

We give an overview in Table \ref{table:no-hwp} where 
we show the probabilities that what Alice ($j=1,2,3,4$) prepares
will be received by Bob ($m=1,2,3,4$) weighted with Eve's
($k=1,2,3,4$) presence $X$, in the right-hand part entitled Bob
as well as Eve's probabilities of gaining Alice's messages again
weighted with her presence $X$, in the left-hand part entitled Eve.

\begin{table}[ht]
\caption{Bob: Probabilities that Bob ($m$) will measure what Alice ($j$)
sends when HWPs are not inserted, weighted by Eve's ($k$) presence
$X$; Eve: Probabilities that Eve will read messages that Alice sends
and Bob receives weighted by her presence $X$; When Eve is not in the
line we have $X=0$ and when she is in the line all the time we have
$X=1$; Numbers 1,\dots,4 in the top two rows and the  column on the
left hand side denote $|\chi^1\rangle,\dots,|\chi^4\rangle$.}
\center
\setlength{\tabcolsep}{6pt}
\begin{tabular}{|c|c|c|c|cc|cc||c|c|cc|cc|c|}
\hline
\multicolumn{2}{|c|}{}&\multicolumn{6}{|c||}{Eve}&\multicolumn{6}{|c|}{Bob}\\
\hline
\multicolumn{2}{|c|}{$j$}
&{1}&{2}&\multicolumn{2}{|c|}{3}&\multicolumn{2}{|c||}{4}&{1}&{2}&\multicolumn{2}{|c|}{3}&\multicolumn{2}{|c|}{4}\\
\hline
\multicolumn{2}{|c|}{$k$}&1&2&3&4&3&4&1&2&3&4&3&4\\
\cline{1-2}
\multirow{4}*{$m$}
&1&$\frac{X}{4}$&&&&&&$\frac{1}{4}$&&&&&\\
\cline{2-2}
&2&&$\frac{X}{4}$&&&&&&$\frac{1}{4}$&&&&\\
\cline{2-2}
&3&&&$\frac{X}{8}$&&$\frac{X}{8}$&&&&$\frac{2-X}{8}$&&$\frac{X}{8}$&\\
\cline{2-2}
&4&&&&$\frac{X}{8}$&&$\frac{X}{8}$&&&&$\frac{X}{8}$&&$\frac{2-X}{8}$\\
\hline
\end{tabular}
\label{table:no-hwp}
\end{table}

When Alice and Bob are in the same basis but with 
both HWPs inserted and Alice sends $|\chi^1\rangle$  
or $|\chi^2\rangle$ her HWP($\frac{\pi}{8}$) 
can be regarded as an operator acting on the states as follows
\begin{eqnarray}
{\rm HWP}|\chi^{1,2}\rangle=\frac{1}{2}
\left[|HH\rangle\mp|VV\rangle
  -(|HV\rangle\pm|VH\rangle)\right]=\frac{1}{\sqrt{2}}
(|\chi^{3,4}\rangle-|\chi^{2,1}\rangle).
\label{eq:alice-hwp}
\end{eqnarray}

At Eve's BS Alice's $|\chi^1\rangle$ will be transformed into the 
following one: 
\begin{eqnarray}
\hskip-10pt\frac{1}{2\sqrt{2}}(|H\rangle_1|H\rangle_1-|V\rangle_1|V\rangle_1
-|H\rangle_2|H\rangle_2+|V\rangle_2|V\rangle_2)
-\frac{1}{2}(|H\rangle_1|V\rangle_1-|H\rangle_2|V\rangle_2)\nonumber\\
=
\frac{1}{2}(|\Phi^-\rangle_{11}-|\Phi^-\rangle_{22})
-\frac{1}{\sqrt{2}}|\chi^2_-\rangle,\!\!\!\!\!\!\!\!\!\!
\label{eq:eve-bs1a}
\end{eqnarray}
where the indices refer to the sides of BS. The states
$|\chi^{2,3,4}\rangle$ transform similarly. 

The probabilities of Eve reading a particular result for the 
states Alice sent will be
\begin{eqnarray}
\!\!\!\!\!\!\!\!\!\!\!|\chi^{1,2}\rangle\longrightarrow|\chi^{2,1}\rangle:\ 50\%;
\ |\chi^3\rangle:\ 25\%; 
\ |\chi^4\rangle:\ 25\%.
\label{eq:eve-read1}
\end{eqnarray}

Alice's ${\rm HWP}(\frac{\pi}{8})$ changes 
$|\chi^{3,4}\rangle$ in a similar way and we obtain 
for the probabilities of Eve's reading:
\begin{eqnarray}
\!\!\!\!\!\!\!\!\!\!\!|\chi^{3,4}\rangle\longrightarrow|\chi^{1}\rangle:\ 25\%;
\ |\chi^2\rangle:\ 25\%; 
\ |\chi^{3,4}\rangle:\ 50\%.
\label{eq:eve-read2}
\end{eqnarray}

In both cases Eve can resend $|\chi^{3,4}\rangle$ only randomly
so that the probabilities for sending states in the latter case 
will be: 
\begin{eqnarray}
\hskip-23pt\ |\chi^1\rangle:\ 25\%;
\ |\chi^2\rangle:\ 25\%; 
\ |\chi^3\rangle:\ 25\%; 
\ |\chi^4\rangle:\ 25\%,
\label{eq:eve-send1}
\end{eqnarray}
i.e., her $|\chi^3\rangle$ resending will be
indistinguishable from her $|\chi^4\rangle$ resending.
In Table \ref{table:yes-hwp} we give the probabilities for
respective measurements with the HWPs inserted for $|\chi^1\rangle$
and $|\chi^3\rangle$ (denoted by 1 and 3 in the top row).
For example, the 2nd probability $X$/16 is the probability that
Bob will measure $|\chi^1\rangle$ when Alice prepared 
$|\chi^1\rangle>$ and Eve measured it as $|\chi^2\rangle$ and encoded
it into a pair which Bob will measure as $|\chi^1\rangle$ due to his 
HWP; the last probability $X$/64 is the probability that Bob will 
measure $|\chi^4\rangle$ when Alice prepared $|\chi^3\rangle$ and Eve
measured it as $|\chi^3\rangle$, but failed to encode it and encoded 
$|\chi^4\rangle$ instead. Double entries divided by a diagonal line 
refer to Eve's and Bob's reading, respectively. For example the 
entries in the first line, first column means Alice sent 
$|\chi^1\rangle$ and Bob will measure $|\chi^1\rangle$ with the 
probability ($X$-1)/4 (i.e., only when Eve is in line, not all the 
time). On the other hand Eve will never measure $|\chi^1\rangle$ in 
this case (0). A table with probabilities for $|\chi^2\rangle$ and
$|\chi^4\rangle$ looks similar and is therefore omitted.

\begin{table}[ht]
\caption{Probabilities that Bob ($m$) will measure what Eve ($k$)
sends him after her reading what Alice ($j$) prepared when HWPs are
inserted. Alice's similar sendings of $|\chi^{2,4}\rangle$ are not 
shown here. Bob-columns and Eve-row contain marginal probabilities.}
\center
\setlength{\tabcolsep}{6pt}
\renewcommand{\arraystretch}{1.5}
\begin{tabular}{|c|c||c|c|c|c|c||c|c|c|c|c|}
\hline
\multicolumn{2}{|c|}{$j$}
&\multicolumn{5}{|c||}{1}
&\multicolumn{5}{|c||}{3}
\\
\hline
\multicolumn{2}{|c|}{$k$}&1&2&3&4&Bob&1&2&3&4&Bob\\
\cline{1-2}\cline{7-7}\cline{12-12}
\multirow{4}*{$m$}
&1&
\mbox{\backslashbox{\lapbox[\width]{-0.1pt}{0}
\kern-1em}{\kern-1em\lapbox[\width]{3pt}{$\frac{1-X}{4}$}}}
&$\frac{X}{16}$&$\frac{X}{64}$&$\frac{X}{64}$&
$\frac{8-5X}{32}$&&$\frac{X}{32}$&$\frac{X}{64}$&$\frac{X}{64}$&$\frac{X}{16}$
\\
\cline{2-2}
&2&&&$\frac{X}{64}$&$\frac{X}{64}$&$\frac{X}{32}$&
$\frac{X}{32}$&&$\frac{X}{64}$&$\frac{X}{64}$&$\frac{X}{16}$
\\
\cline{2-2}
&3&&$\frac{X}{32}$&$\frac{X}{32}$&&$\frac{X}{16}$&$\frac{X}{64}$&$\frac{X}{64}$&
\mbox{\backslashbox{\lapbox[\width]{-2pt}{$\frac{X}{32}$}
\kern-1em}{\kern-1em\lapbox[\width]{3pt}
{$\frac{8-7X}{32}$}}}&&$\frac{4-3X}{16}$
  \\
\cline{2-2}
&4&&$\frac{X}{32}$&&$\frac{X}{32}$&$\frac{X}{16}$&$\frac{X}{64}$&
$\frac{X}{64}$&&$\frac{X}{32}$&$\frac{X}{16}$
\\
\hline
\multicolumn{2}{|c|}{Eve}&0&$\frac{X}{8}$&$\frac{X}{16}$&$\frac{X}{16}$
&\mbox{\backslashbox{$\frac{X}{4}$\kern-1em}{\kern-1em$\ \ \frac{1}{4}$}}&
$\frac{X}{16}$&$\frac{X}{16}$&$\frac{X}{16}$&$\frac{X}{16}$
&\mbox{\backslashbox{$\frac{X}{4}$\kern-1em}{\kern-1em$\ \ \frac{1}{4}$}}\\
\hline 
\end{tabular}
\label{table:yes-hwp}
\end{table}

Let us now calculate the mutual information between Alice 
and Bob. We make use of its standard definition which is
applicable to any communication protocol
\cite[Eq.~(4)]{moroder-lutkenhaus-06}. 
\begin{eqnarray}
I_{AB}:=H(A_B)+H(B)-H(A,B),
\label{eq:mi}
\end{eqnarray}
where the Shannon entropy and the Shannon joint entropy
are defined as \cite[Sec.~II]{moroder-lutkenhaus-06}
\begin{eqnarray}
H(B)&=&-\sum_{i,n}p(a_i,b_n)\log_2p(b_n),\nonumber\\
H(A,B)&=&-\sum_{i,n}p(a_i,b_n)\log_2p(a_i,b_n),
\label{eq:mia}
\end{eqnarray}
where $p(a_i,b_n)$ is the joint probability that $A=a_i$
and $B=b_n$, and $p(b_n)$ are marginal probabilities; 
base 2 of the logarithm gives the entropy and 
information in bits. $H(A_B)$ is the entropy calculated with
respect to the messages received by Bob and therefore we have
$H(A_B)=H(B)=4\frac{1}{4}\log_2\frac{1}{4}=2$ bits. We define
$I_{AE}$ in an analogous way with $H(A_E)=H(E)=2X-X\log_2X$.
We obtain the probabilities $p(a_i,b_n)$ from the probabilities  
$p_{jkm}$ which are given in Tables \ref{table:no-hwp} 
and \ref{table:yes-hwp} ($i$ is $j$ for Alice and $n$ is
$k$ for Eve or $m$ for Bob).

After a somewhat lengthy but straightforward calculation
we obtain the mutual information after sifting with and without
HWPs, the arithmetical means of which are given as the
following functions of $X$: 
\begin{eqnarray}
\hskip-18pt I_{AB}(X)&=&\frac{1}{2}+\frac{X}{8}+\frac{1}{32}
\Big(15X\log_2X+4(2-X)\log_2(2-x)\nonumber\\
&+&2(4-3X)\log_2(4-3x)+(8-5X)\log_2(8-5x)\Big)\nonumber\\
\hskip-18pt I_{AE}(X)&=&\frac{7X}{8}-X\log_2X.
\label{eq:iabe}
\end{eqnarray}

\begin{figure}[hbt!]
\centering
\includegraphics[width=0.5\textwidth]{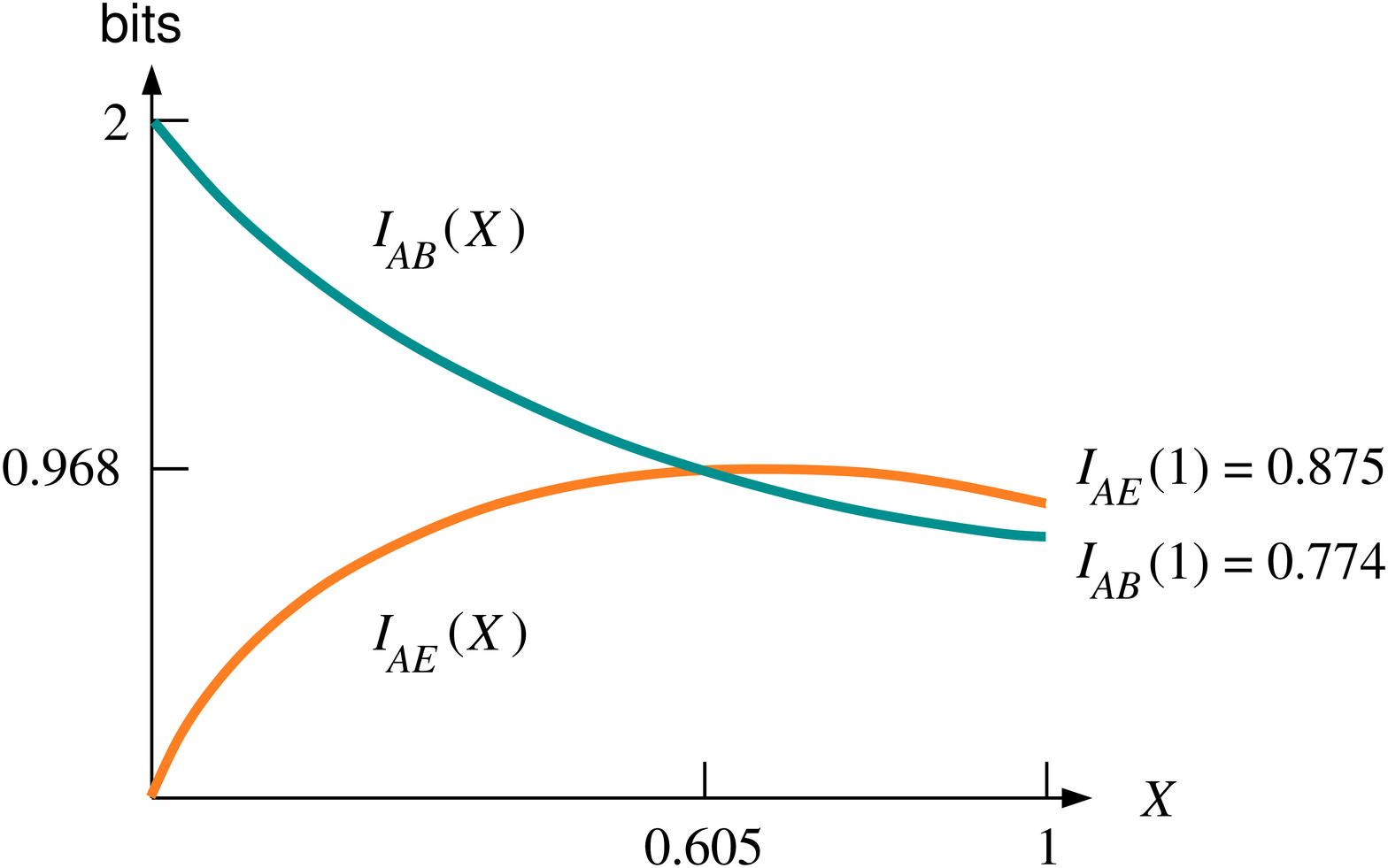}
\caption{\label{fig:pabe}{Mutual information of
Alice and Bob, $I_{AB}(X)$, vs.~the one of Alice and Eve
$I_{AE}(X)$ as functions of Eve's presence, $X$, in the line.}}
\end{figure}

Their plots in Fig.~\ref{fig:pabe} highlight the key result of our 
security analysis. 

When Eve is in the line all the time we have
$I_{AB}(1)=\frac{5}{8}+\frac{3}{32}\log_23=0.774$
and $I_{AE}(1)=\frac{7}{8}=0.875$. Remarkably, the difference 
between $I_{AE}(X)$ and $I_{AB}(X)$ is even higher for some 
values of $0.605<X<1$, for which $I_{AE}(X)>I_{AB}(X)$.

Another method to estimate the security of our protocol is
by means of the control mode. When only Bob's HWP is in place
and Alice sends $|\chi^{3,4}\rangle$, according to
Eq.~(\ref{eq:eve-read2}), Eve will receive them as 
$|\chi^{3,4}\rangle$ and Bob should receive them as
$|\chi^1\rangle$ or $|\chi^2\rangle$ with the probability of
25\%, each, or as $|\chi^{3,4}\rangle$ with the probability of
50\%. However, in half of the cases Eve will fail to resend
the latter states, i.e., she will send them incorrectly as
$|\chi^{4,3}\rangle$ instead, with the probability of 25\% and Bob
will immediately detect Eve's presence due to Alice's classical
information [see Eq.~(\ref{eq:eve-send1})] via Eve's bit-flips
$|\chi^3\rangle\to|\chi^4\rangle$
($|\chi^4\rangle\to|\chi^3\rangle$) with the probability of 1/4,
and therefore Eve's probability of escaping detection during
each of these two sendings is $1-1/4=3/4$.  

Within a complete set of 4 different messages $|\chi^{1-4}\rangle$
in the control mode Eve's probability of avoiding detection with
either $|\chi^1\rangle$ or $|\chi^2\rangle$ is 1 and with
$|\chi^3\rangle$ or $|\chi^4\rangle$ is 3/4. Alice's sendings come
one after another and therefore the probabilities multiply and Eve
will avoid detection within a single cycle with the probability of
$1\times 1\times (3/4)\times (3/4)=(3/4)^2\approx 0.56$. After a
more detailed analysis we arrive at a result that with such
repeated trials Eve's probability of snatching one character 
(8 bits) undetected is $(0.53/1.54)^8\approx 0.0002$. We do not 
have to sacrifice data in order to detect Eve in this way. 

\section{Discussion}
\label{sec:conclusion}

To summarize, we introduced a high capacity (2 bits) protocol that
relies on a mixed state basis consisting of two Bell states and two
states from the computational basis (a kind of blending ping-pong
(pp) and BB84-like protocols) which can be realized experimentally
right away since it relies only on off-the-shelf components. The
protocol is supported by classical information exchanged between
Alice and Bob over a classical channel as shown in
Fig.~\ref{fig:bs-bs}. When both HWPs are inserted or none, photon
states are in the same bases and the messages are being transferred.

When they are in different bases (only one of the HWPs is inserted)
Alice and Bob will detect Eve's bit-flips with the probability of
99.98\%\ during her snatching of her first byte of messages as shown
in Sec.~\ref{sec:security}. So, the different bases do not only
support the transfer of messages but function as a control mode as
well, similarly to such a mode in the pp protocol and
contrary to BB84-like protocols where different bases transmissions
are simply discarded (and a portion of messages must be sacrificed
for QBER verification). 

Still, Eve can hide behind the exponential losses in the fibers
and we carried out a security analysis in Sec.~\ref{sec:security}
to estimate at which level of Eve's presence Alice and Bob must
abandon the transmission for the chosen attack. The attack we
chose to consider is a modification of Nguyen's attack
\cite{nguyen-04} shown in Fig.~\ref{fig:bs-eve-bs}. When applied 
to the standard pp-like protocols it can be viewed as sending plain 
text messages protected by the control mode. For a modified 
two-state pp protocol with a vacuum state it can be proved secure 
\cite{han-14,chen-16}, but for the standard pp protocol or its 
extension to four states there is no critical presence of Eve in 
the protocol since we have constant and maximal Alice-Bob mutual 
information ($I_{AB}$) for 
any Eve's presence ($0<X<1$) and without it and without having a 
new kind of privacy amplification algorithms developed for absent 
critical presence (disturbance, QBER) we do not know when to abort 
the transmission in such a protocol. In contrast, our protocol is 
resistant against such Nguyen's attack because it also contains 
entanglement-based computational basis states.

On the other hand, it is also fundamentally different from the
BB84 because Eve cannot send her photon particularly polarized
without also affecting Bob's photon's polarization, i.e., she
cannot deterministically resend photons in a particular state of
polarization even when she knows whether HWPs are inserted of not.

A modified pp protocol with a vacuum state proposed in 
\cite{han-14,chen-16} proved to be secure. In other  
pp-like protocols
\cite{dusek-lutkenhaus-06,wojcik-03,bostrom-felbinger-08,pavicic-pra-13,shapiro-06,shapiro-wong-06,shapiro-10,niederberger-05}, whenever one 
can define a critical disturbance, Eve's attacks influence $I_{AB}$ 
with respect to $I_{AE}$ more than in our protocol. As shown in 
Fig.~\ref{fig:pabe}, for Eve's presence of up to 60\%\ ($X<0.605$) 
we have $I_{AB}>I_{AE}$ and the transfer is secure for the considered
attack. This Eve's presence corresponds to the disturbance of 30\%\ 
($D=X/2$) which is much higher than 11\%\ and 12.7\%\ of $D$ (QBER) 
for four- and six-state BB84 protocol and also higher than 
22.7\%\ and 25\%\ for the 3- and 4-dim protocols mentioned in 
Sec.~\ref{sec:intro}. 

Recently, two-way classical communication channel was used to 
boost the critical QBER of four- and six-state BB84 protocols to 
26\%\ and 30\%, respectively \cite{wang-04}. Similar two-way 
classical communication channel can be used to boost our critical 
QBER significantly over 30\%. This is the work in progress.

Taken together, the proposed protocol allows for much higher
disturbance (QBER, Eve's presence), at which the mutual information
between Alice and Eve reaches the mutual information between Alice 
and Bob, than other standard pp-like protocols. The price we have 
to pay for such an increased robustness of the protocol is a limited 
distance since the efficiency of Bob detecting both photons 
diminishes over four times the distance that a single photon would 
cover in a BB84 implementation. Hence, right now, the protocol is 
suitable for urban inter-institutional high-security networks.

\section*{Funding}
Financial supports by the DFG (Sfb787), BMBF (Q.com-H), and 
EMPIR 14IND05 MIQC2 are acknowledged. M.P.~acknowledges funding 
by the Alexander von Humboldt Foundation, the Croatian Science 
Foundation through project IP-2014-09-7515, and the Ministry of 
Science and Education of Croatia through CEMS.
J.W.~acknowledges funding by the EU project 702304-3-5-FIRST and
the Humboldt Graduate School (PostDoc Scholarship). 

\section*{Acknowledgment.} 
We thank Alejandro Saenz for discussions and valuable comments. 

\end{document}